\documentclass[pre,twocolumn,tightenlines,superscriptaddress]{revtex4-1}

\usepackage[utf8]{inputenc}

\usepackage[T1]{fontenc}
\usepackage{bbold}
\usepackage{amssymb}
\usepackage{amsmath}
\usepackage{graphicx}
\usepackage{color}
\usepackage[colorlinks=true,linkcolor=blue,citecolor=red,urlcolor=blue]{hyperref}
\usepackage[section]{placeins}

\usepackage{enumitem,amssymb}
\newlist{todolist}{itemize}{2}
\setlist[todolist]{label=$\square$}
\usepackage{pifont}

\newcommand{\be}{\begin{equation}}
	\newcommand{\ee}{\end{equation}}

\def\i{\text{i}}
\newcommand{\bra}[1]{\langle{#1}|}
\newcommand{\ket}[1]{|{#1}\rangle}
\newcommand{\bkt}[2]{\langle{#1}|{#2}\rangle}

\newcommand{\floor}[1]{\lfloor #1 \rfloor}

\usepackage{ulem}

\begin{document}
	
	\title{Anderson localization of a Rydberg electron}
	
	\date{\today}

	\author{Matthew T. Eiles}
	\email{meiles@pks.mpg.de}
	\affiliation{Max-Planck-Institut f\"ur Physik komplexer Systeme, N\"othnitzer Str.\ 38, 
		D-01187 Dresden, Germany }

	\author{Alexander Eisfeld}
	\affiliation{Max-Planck-Institut f\"ur Physik komplexer Systeme, N\"othnitzer Str.\ 38,
		D-01187 Dresden, Germany }

	\author{Jan M. Rost}
	\affiliation{Max-Planck-Institut f\"ur Physik komplexer Systeme, N\"othnitzer Str.\ 38,
		D-01187 Dresden, Germany }

	\begin{abstract}
		Highly excited Rydberg atoms inherit their level structure, symmetries, and scaling behavior from the hydrogen atom. 
		We demonstrate that these fundamental properties enable a thermodynamic limit of a single Rydberg atom subjected to interactions with nearby ground state atoms.
		The 
		limit is reached by simultaneously increasing the number of ground state atoms and the level of excitation of the Rydberg atom, for which the Coulomb potential supplies infinitely many and highly degenerate excited states. 
		Our study reveals a surprising connection to an archetypal concept of condensed matter physics, Anderson localization, facilitated by a direct mapping between the Rydberg atom's electronic spectrum and the spectrum of a tight-binding Hamiltonian.
		The hopping amplitudes of this tight-binding system are determined by the arrangement of ground state atoms and can range from oscillatory and long-ranged to nearest-neighbor. In the latter we identify clear signatures of the Anderson localization of the Rydberg electron.
	\end{abstract}

		\maketitle

\section{Introduction}
		The origin of quantum mechanics is inextricably linked to the bound state spectrum of hydrogen, which consists of an infinite series of discrete levels labeled by an integer-valued principal quantum number $\nu$  \cite{bohrConstitution1913,pauliUeber1926,schrodingerUndulatory1926}.
		Because of hydrogen's underlying $SO(4)$ symmetry, these levels are $\nu^2$-fold degenerate \cite{ banderGroup1966,gallagherRydberg2005}.  
		This enhances the effect of external perturbations, as evinced by the response of hydrogen atoms to electric and magnetic fields \cite{friedrichHydrogen1989a} or to electron scattering \cite{gailitisInfluence1963,sadeghpourDominant1990}. 
		The study of these aspects exposes deep connections between the {excited} electronic structure of hydrogen and seemingly disparate physical arenas. 
		Compelling examples include the hydrogen atom in a strong magnetic field, which is fundamental to quantum chaos and non-linear dynamics \cite{delandeQuantum1986,wintgenIrregular1989}, and the organization of doubly-excited H$^-$ states into multiplets, a phenomenon akin to the symmetry classifications ubiquitous in elementary particle physics \cite{tannerTheory2000a}.
		{In this article}, we forge a connection between
		the hydrogen atom and 
		condensed matter via Anderson localization.  

  {Hydrogen's properties are shared by the highly excited \textit{Rydberg states} of other atomic species, since the influence of the multielectron core essentially vanishes for these exaggerated states characterized by large $\nu$ values, almost millisecond lifetimes, and micron-scale orbits \cite{seatonQuantum1983,gallagherRydberg2005}. 
Localized perturbations to a Rydberg atom, caused by the scattering of its electron off of one or more ground state atoms -- denoted scatterers in the following -- mix the degenerate states within each $\nu$ manifold, giving the otherwise weak interaction of the scatterers a surprisingly strong effect \cite{greeneCreation2000,shafferUltracold2018a}.
Recently, optical tweezer arrays have become available which can hold ground state atoms in nearly arbitrary arrangements \cite{bernienProbing2017,browaeysManybody2020,bluvsteinControlling2021a}.
This allows for the possibility to create a \textit{Rydberg composite} by perturbing a Rydberg atom with a predetermined configuration of point-like impurities \cite{hunterRydberg2020a}. }

\begin{figure*}[t]
			\includegraphics[width=0.9\textwidth]{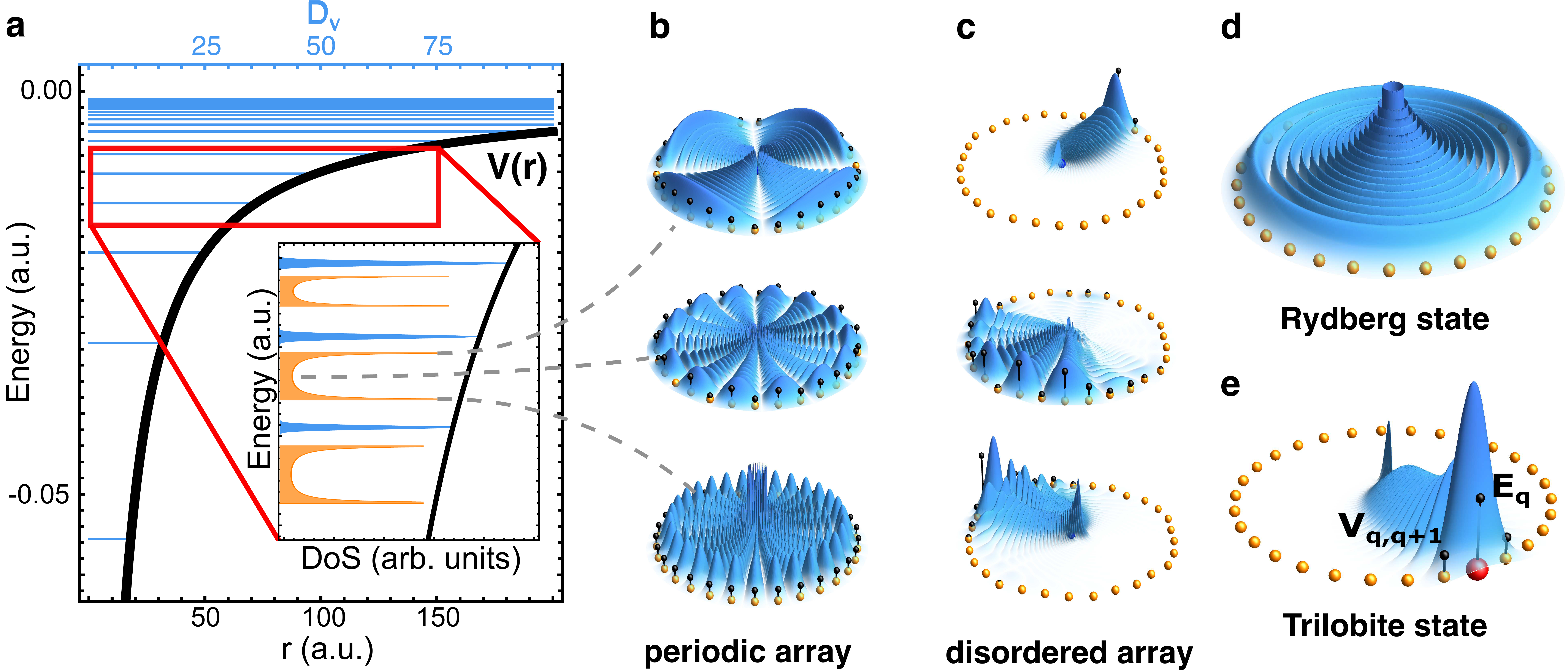}
			\caption{\label{fig:intro} 
				(a)  The level structure of the Rydberg composite. The Coulomb potential $V(r)=-1/r$ supports an infinite bound spectrum,  $E_\nu=-1/2\nu^2$, denoted with blue lines. The length of each line represents the level degeneracy $D_\nu=\nu^2$ and the typical size of the electronic states, $\langle r\rangle\sim \nu^2$. 
				The inset highlights the densities of states (DoS) of three Rydberg levels when the atom is perturbed by a ring of $M$ scatterers with radius $2\nu^2$. 
				A highly structured DoS consisting of $M$ perturbed states (orange) forms, shifted away from the unaffected $M - \nu^2$ states (blue).
				In the thermodynamic limit, $M,\nu\to\infty$, the bandwidth and center of mass of the shifted DoS are (within an overall scaling factor) independent of $M$ and $\nu$.
			(b,c) The eigenstate amplitudes located at the marked positions in the DoS for both periodic and disordered arrays are shown in the Rydberg (blue) and site (black spheres) representations.  
				Both representations exhibit the same features in the vicinity of the scatterers (orange spheres).
				(d) An exemplary Rydberg state, which is spherically symmetric and delocalized. 
			(e) An exemplary trilobite state for the scatterer $q$ marked in red. The trilobite's amplitude at $q$ determines the on-site potential $E_q$, while its amplitude at ${q'}$ determines  the hopping amplitude $V_{qq'}$. 
			}
		\end{figure*}

		Fig.~\ref{fig:intro}(a)  illustrates the level structure of such a Rydberg composite, formed after the immersion of $M$ scatterers within the Rydberg wave function. 
		Many states in each $\nu$ manifold are not affected, but a subspace of dimension $M$ splits away and possesses a density of states which depends non-trivially on the scatterer arrangement \cite{hunterRydberg2020a}. The spectrum of this perturbed subspace coincides identically with that of a tight-binding Hamiltonian \cite{eilesRing2020}
		\begin{equation}
			\label{eq:tightbindinghamiltonian}
			H = \sum_{q=1}^ME_q\ket{q}\bra{q} + \sum_{q=1}^M\sum_{q'\ne q}^MV_{qq'}\ket{q}\bra{q'},
		\end{equation}
		where $\{\ket{q}\}$ is a basis of states localized on individual sites.
		The on-site potentials $E_q$ and hopping amplitudes $V_{qq'}$ arise from the Rydberg electron's motion in the confluence of the infinite-ranged Coulomb and zero-ranged electron-scatterer potentials. 
		Eq.~\ref{eq:tightbindinghamiltonian} creates an unexpected conceptual link between a Rydberg atom interacting with many ground state atoms and the dynamics of a particle hopping through a lattice. 
		
  We exploit this {link} to demonstrate that a Rydberg electron can undergo Anderson  localization: in the thermodynamic limit of infinite system size, the entire spectrum of electron eigenstates exponentially localizes in the presence of arbitrarily weak disorder \cite{andersonAbsence1958,thoulessElectrons1974,leeAnderson1981,abrahamsScaling1979,eversAnderson2008}.
		To construct the thermodynamic limit in the Rydberg system we 
		determine a relationship between $M$ and $\nu$ such that increasing them in tandem -- relying on the infinite series and scaling relations of Rydberg levels  --  
		leads to a well-defined Hamiltonian whose matrix elements are independent of its size. 
		We study effectively one-dimensional localization by placing the scatterers on a ring around the Rydberg atom's core, and then randomly disordering their positions. 
		Different ring radii lead to different hopping amplitudes, ranging from the nearest-neighbor interactions conventionally studied to more unusual long-range and sign-changing interactions.
		This flexibility {gives rise to} 
		a variety of Anderson models. 
		
	\section{Mapping Rydberg dynamics to a lattice model}	
		The bare Rydberg states $\ket{\nu i}$ are labeled by the principal quantum number $\nu$ and a collective index $i = \{l,m\}$ for the angular momentum quantum numbers, $0\le l \le \nu-1 $ and $|m|\le l$. 
		We consider each $\nu$ manifold individually since the perturbation from the $M$ scatterers is too weak to couple different manifolds. 
		In each of the resulting $D_\nu$-dimensional degenerate subspaces, where $D_\nu = \nu^2$, the Hamiltonian matrix elements are
		\begin{equation}
			\label{eq:originalham}
			\mathcal{H}_{ii'} = -\frac{1}{2\nu^2}\delta_{ii'}+2\pi\sum_{q=1}^Ma_s[k(R_q)]\bkt{\nu i}{\vec R_q}\bkt{\vec R_{q}}{\nu i'}
		\end{equation}
		in atomic units. 
{The first term is the Rydberg atom's energy, and the second is a sum over zero-range pseudopotentials describing the electron-scatterer interaction in terms of the $s$-wave electron atom scattering lengths $a_s[k(R_q)]$ and the amplitudes $\bkt{\nu i}{R_q}$ of the Rydberg states at the scatterer positions} \cite{fermiSopra2008,greeneCreation2000}.  Appendix \ref{sec:app:hamiltonian} provides further background for Eq.~\ref{eq:originalham}, including a discussion of its generality.

		Expressing $\mathcal{H}$ in terms of the rectangular matrix $\mathcal{W}_{iq}=\sqrt{a_s[k(R_q)]}\bkt{\nu i}{\vec R_q}$ makes explicit its separable form and shows that $\text{rank}(\mathcal{H})= M$.  We project onto the image of $\mathcal{H}$ using $U = (\mathcal{W}^\dagger\mathcal{W})^{-1/2}\mathcal{W}^\dagger$, which is a \textit{semi-unitary} transformation since $U$ and its Moore-Penrose right inverse $U^\dagger$ satisfy $UU^\dagger = \mathbb{1}_M$ and $U^\dagger U\ne \mathbb{1}_{D_\nu}$. However, we can still transform  $\mathcal{H}$  into $ H =U\mathcal{H}U^\dagger$ using $U^\dagger U\mathcal{H}U^\dagger U=\mathcal{H}$, where
		\begin{equation}
			\label{eq:compositeham}
			H=\sum_{q,q'=1}^M\ket{q}\left( -\frac{1}{2\nu^2}\delta_{qq'}+2\pi\sum_{i=1}^{D_\nu}\mathcal{W}_{qi}^\dagger\mathcal{W}_{iq'}\right)\bra{q'}
		\end{equation}
		is a tight-binding Hamiltonian in the form of Eq.~\ref{eq:tightbindinghamiltonian} and possessing the same non-zero eigenenergies as $\mathcal{H}$.
		Its eigenvectors $\ket{\Psi_k} = \sum_{q = 1}^Mc_q^{(k)}\ket{q}$ transform back into the Rydberg basis via $\ket{\Psi_k}_\text{Ryd} = \sqrt{2\pi/E_k}\sum_{i=1}^{D_\nu}\mathcal{W}_{iq} c_q^{(k)}\ket{\nu i}$, where $E_k$ is the eigenenergy.
{  Appendix \ref{sec:app:transform} describes this transformation in more detail.}
		Fig.~\ref{fig:intro}(b) and (c) display exemplary eigenstates in both representations for both a periodic and a disordered scatterer array. 
		
		Eq.~\ref{eq:compositeham} reveals the connection between the perturbed Rydberg spectrum and a tight-binding Hamiltonian. 
		A physical interpretation emerges upon considering the so-called ``trilobite{''} eigenstate of a Rydberg atom and a single scatterer, $\ket{T_q} = \sum_{i=1}^{D_\nu}\mathcal{W}_{iq} \ket{\nu i}$ \cite{greeneCreation2000,boothProduction2015a}.
		Unlike the spherically symmetric  eigenstates of the bare Rydberg atom, which 
		extend over the entire scatterer array (Fig~\ref{fig:intro}(d)), the trilobite state  is peaked at the scatterer's position (Fig.~\ref{fig:intro}(e)).
		The matrix elements $H_{qq'}$ are proportional to the trilobite overlaps $\bkt{T_q}{T_{q'}}$ or, equivalently, the amplitudes $\bkt{\vec R_{q'}}{T_q}$ \cite{liuPolyatomic2006,eilesUltracold2016,eilesTrilobites2019}.
		This provides a convenient means to pictorially estimate the properties of the tight-binding Hamiltonian, as in Fig.~\ref{fig:intro}(e).
	Futhermore,	closed-form expressions for $\bkt{T_q}{T_{q'}}$ 
		simplify calculations and facilitate asymptotic expansions, as discussed in Appendix \ref{sec:app:matrixel} \cite{eilesUltracold2016,eilesTrilobites2019}.

	\section{Thermodynamic limit}	
In a typical solid-state system described by a tight-binding Hamiltonian Eq.~\ref{eq:tightbindinghamiltonian}, the elements $E_q$ and $V_{qq'}$ are independent of $M$ and the thermodynamic limit is reached by increasing the system's size, i.e.\  $M\to\infty$. 
However, the matrix elements $H_{qq'}$ of Eq.~\ref{eq:compositeham} depend strongly on both $\nu$ and $M$: the Rydberg atom's size and energy scales are  $\nu$-dependent, and the hopping amplitudes depend on the distance, inversely proportional to $M$, between scatterers. 
As an initial step in separating these scales, we accomodate the overall size of the Rydberg wave function by parameterizing the ring's radius {as} 
$2\nu^2R$, where $R\in [0,1]$.  {This parametrization ensures that systems with different $\nu$ but identical $R$ values 
have similar properties \cite{eilesRing2020}, and the range of $R$ keeps} 
the scatterers 
within the classically allowed region. 
We {will} discuss three specific cases in detail in this article: $R = 1$, $R = 0.75$, and $R = 0.5$. 

In a subsequent step, for each $R$ we eliminate the $M$-dependence at a coarse-graining level by fixing $M$ as a function of $\nu$ such that the inter-scatterer distance, and hence the hopping amplitudes, are invariant with respect to changes in $\nu$. 
 The functional form of $M(\nu)$ hinges on the resolving power of the Rydberg wave functions.
A useful heuristic is that Rydberg states can resolve as many in-plane scatterers as they have available azimuthal nodes, requiring a linear relationship $M(\nu)=\nu$ for most $R$ values. 
{For $R\to 1$ 
those} Rydberg states possessing 
 the many azimuthal nodes needed to resolve scatterers become exponentially small. 
Thus, fewer scatterers can be resolved and a sublinear relationship is required. 
In particular, for the case $R = 1$, we set $M(\nu)\sim\nu^{2/3}$ (specifically,  $M=\floor{3\nu^{2/3})}$, where $\floor{x}$ is the integer part of $x$).
For the cases $R=0.5$ and $R=0.75$ we use the linear scaling $M(\nu)=\nu$. 
We then extract the residual $\nu$-dependence of the matrix elements $H_{qq'}$. 
 For $R = 1$ we find that the matrix elements are proportional to $\nu^{-13/3}$, but for $R = 0.75$ they are proportional to $\nu^{-4}$. 
The matrix elements of the $R = 0.5$ case do not simultaneously possess a global $\nu$-dependence. 
All of these scaling laws are discussed in further detail below and in Appendix \ref{sec:app:matrixel}.

 \begin{figure}[t]
		
   \includegraphics[width=8.7cm]{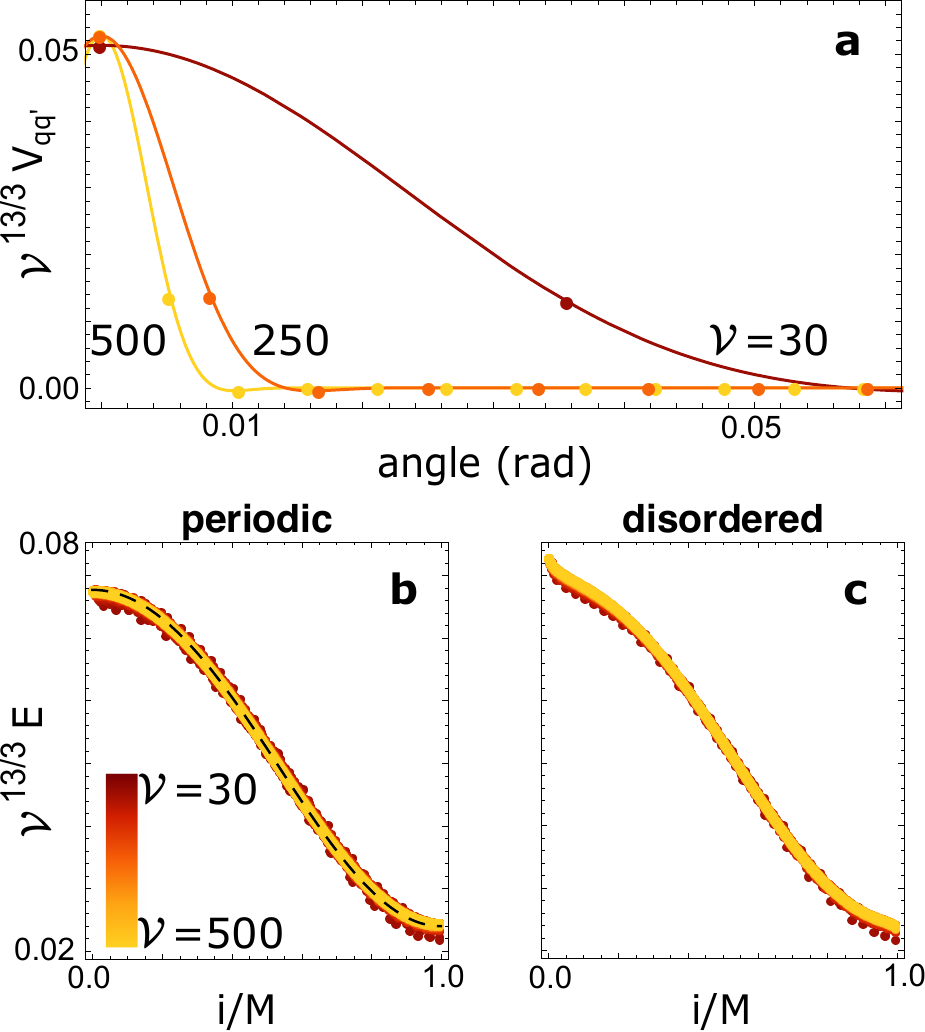}
 	\caption{\label{fig:energies} Characteristic energies and scaling laws for  \textit{R} = 1.  (a) Hopping amplitudes as a function of angle around the ring. The angular positions of the scatterers are marked with points. (b),(c)  Dispersion relations for $30\le \nu\le 500$ in increments of 5. As $\nu$ increases these discrete spectra tend towards the continuous analytic dispersion relation obtained from a model Hamiltonian for $R = 1$ discussed in Appendix \ref{sec:app:matrixel}, shown as the dashed black curve in (b). }
		\end{figure}

Now, we are in a position to factor out an overall $\nu$-dependence such that the matrix elements $H_{qq'}$, for fixed $R$ and $M(\nu)$, are independent of $\nu$. 
Taking advantage of the infinite series of Rydberg levels, 
the thermodynamic limit of a Rydberg atom {is realized with} $\nu\to\infty$.
Fig.~\ref{fig:energies}(a) illustrates this analysis for the $R = 1$ case. 
The three curves show the angular dependence of the trilobite states with $\nu=30$, $250$, and $500$.
When appropriately scaled, these functions have identical values when evaluated at the site positions, and hence the matrix elements of $H$ become independent of $\nu$ and $M$. 
As can also be seen in the full trilobite picture in Fig.~\ref{fig:intro}(e), the hopping amplitudes here are non-negligible only for the nearest neighbor site. 
The scaled eigenspectra, shown in 
Fig.~\ref{fig:energies}(b) for $\nu\in[30,500]$, are likewise independent of $\nu$.

  \subsection{Disorder and numerical methods}
{We introduce disorder by randomly varying the positions of the scatterers, either shifting them radially off of the ring or perturbing the angles between them.
While angle disorder results in anti-correlated off-diagonal disorder in the matrix elements $H_{qq'}$, radial disorder leads to uncorrelated on-site disorder and correlated off-diagonal disorder in $H_{qq'}$.
The disorder scaling requires additional analysis since it is not clear \textit{a priori} that the disorder in position has the same $\nu$-dependence as the resulting disorder in the matrix elements. 
For example, although angular disorder leads to first-order energy disorder shifts with the same $\nu$-scaling for all considered $R$ values, in the $R = 1$ and $R = 0.5$ cases radial disorder leads to additional $\nu$-dependencies that must be removed by scaling the positional disorder with $\nu$. 
For $R = 1$ the radial disorder strength must be diminished as $\nu^{-2/3}$. 
These details are discussed further in Appendix \ref{sec:app:matrixel}. 
 Fig.~\ref{fig:energies}(c) shows exemplary eigenspectra for weak radial disorder in the $R=1$ ring.}
We used exact diagonalization to obtain the Rydberg composite's eigenspectrum, averaging over $\mathcal{N}=1000$ disorder realizations. 

		An accurate extrapolation to the thermodynamic limit demands the study of very high $\nu$. 
		The transformed Hamiltonian (Eq. \ref{eq:compositeham}) provides a clear numerical advantage over brute-force diagonalization of the Rydberg Hamiltonian (Eq.~\ref{eq:originalham}) due to the reduced matrix dimension.
		For the largest $\nu$ studied here, $500$, we diagonalize a matrix of dimension $500$ in the site representation; this (fully dense) matrix has dimension $2.5\times 10^5$ in the Rydberg representation.

\subsection{Localization measures}
To quantify the extent of localization and systematically show that all eigenstates localize in the thermodynamic limit, one typically examines statistical properties of the eigenspectrum \cite{oganesyanLocalization2007,shklovskiiStatistics1993} or, as we do here, the eigenstates directly \cite{kramerLocalization1993,mirlinStatistics2000}.
The normalized participation ratio, defined for the eigenstate $\ket{\Psi_k}= \sum_{q=1}^M c_q^{(k)}\ket{q}$ as 
\begin{equation}
\label{eq:NPRk}
\mathcal{P}(k)=\left(M\sum_{q=1}^M| c_q^{(k)}|^4\right)^{-1},
\end{equation}
is a good indicator of the localization length. 
In a maximally localized (delocalized) state, $\mathcal{P}\to 1/M$ ($\mathcal{P}\to1$).
Perfectly delocalized states with strictly real coefficients are characterized by $\mathcal{P}(k) = 2/3$, and therefore we consider states with $\mathcal{P}\ge 2/3$ to be extended. 
In Appendix \ref{sec:app:loca} we demonstrate that the participation ratio in the site basis is equivalent to a spatial participation ratio measured at the scatterer positions. 
Localization therefore occurs simultaneously in both representations.

\section{Examples of different types of interactions}
\subsection{\textit{R}=1: nearest-neighbor interactions}

\begin{figure*}[t]
	\includegraphics[width=\textwidth]{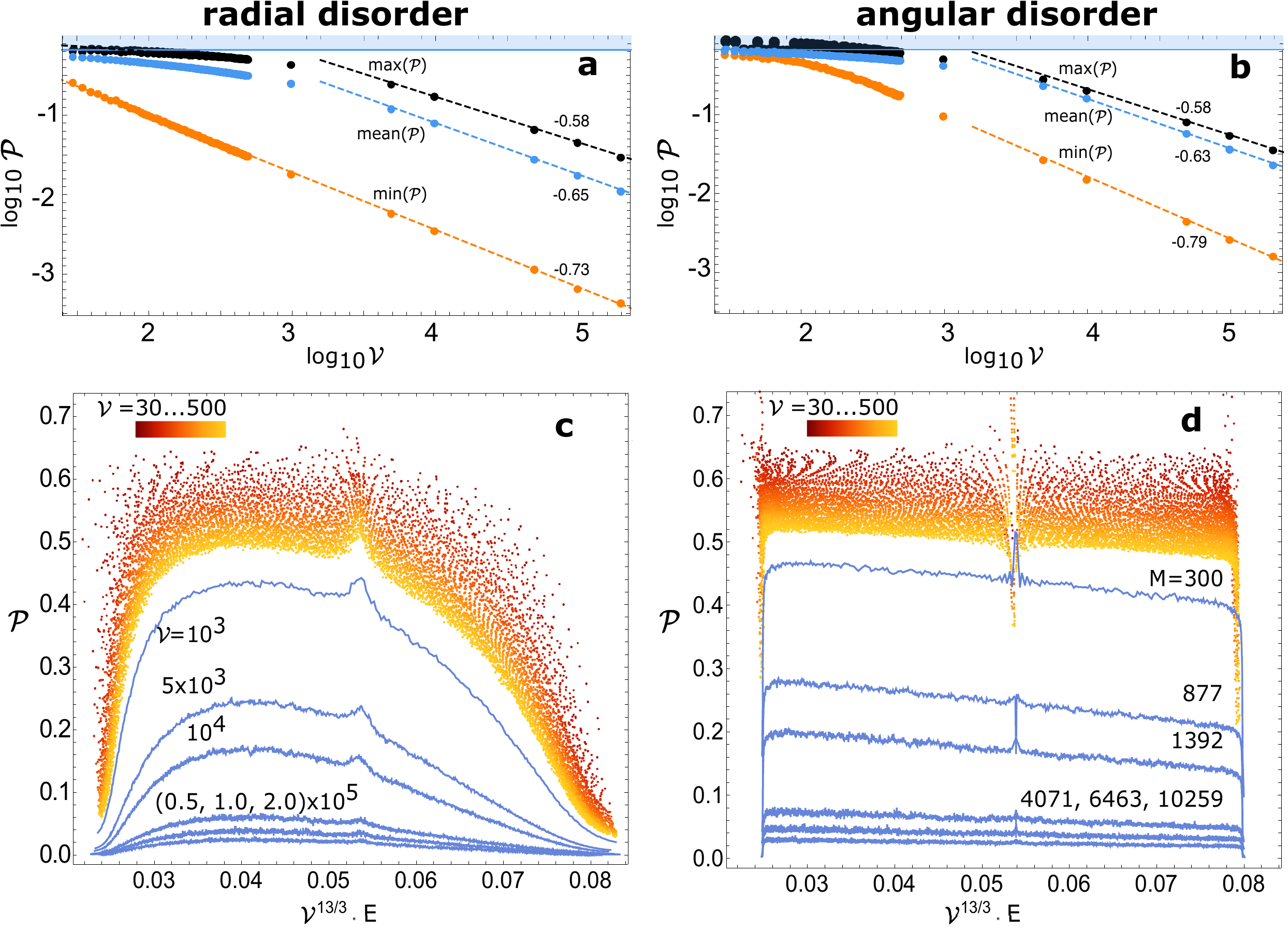}
	\caption{\label{fig:R1} (a,b) the minimum, mean, and maximum values of the normalized participation ratios for radial and angular disorder, respectively. The dashed lines show the asymptotic behavior $\mathcal{P}\sim \nu^\gamma$, labeled by the numerical fit values for $\gamma$.  
		(c,d), the energy-resolved distributions for $30\le\nu\le 500$, using the exact Hamiltonian, and $10^3\le\nu\le10^5$ (blue curves) using the asymptotic model Hamiltonian. Note that the equivalent $M$ values are used as labels in ({d}). }
\end{figure*}

As implied by the nearest-neighbor hopping terms revealed by Fig.~\ref{fig:energies}, the $R = 1$ case allows for a direct comparison with the standard Anderson model. 
The key results for this $R$ value are displayed in Fig.~\ref{fig:R1}. 
In  Fig.~\ref{fig:R1} (a) and (b) we characterize the extent of localization by plotting the minimum, mean, and maximum values of $\mathcal{P}$ as a function of $\nu$.
The fixed disorder strength is sufficiently weak such that extended states having $\mathcal{P}>2/3$ are still present for the lowest $\nu$ values. 
Numerical power-law fits of this data show that $\langle\mathcal{P}\rangle\sim \nu^{-2/3}\sim M^{-1}$, where $\langle\rangle$ denotes an average over the entire spectrum and disorder realizations. 
This numerical evidence clearly indicates that all eigenstates of the $R = 1$ ring localize in the thermodynamic limit.

To obtain these quantitative results, we performed calculations for $\nu\gg 500$, where the exact Hamiltonian becomes numerically cumbersome to evaluate. 
For these values, we used the model Hamiltonian containing only nearest and next-nearest neighbor hopping amplitudes detailed in Appendix \ref{sec:app:matrixel}. 
These amplitudes, obtained asymptotically as $\nu\to\infty$, give a quantitatively accurate model even for relatively low $\nu$ values $\nu\sim100$. 
We demonstrate in Fig.~\ref{fig:energies}(b) that the 
 spectra of the exact system and this model agree excellently for all $\nu$ values, and we then used the $R = 1$ model Hamiltonian for $\nu>500$.
 Here, a calculation in the full Rydberg basis would have involved the diagonalization of a dense matrix of dimension $10^{11}$ rather than the sparse matrix of dimension $10^5$ in the site representation.

\begin{figure*}[t]
	\includegraphics[width=0.95\textwidth]{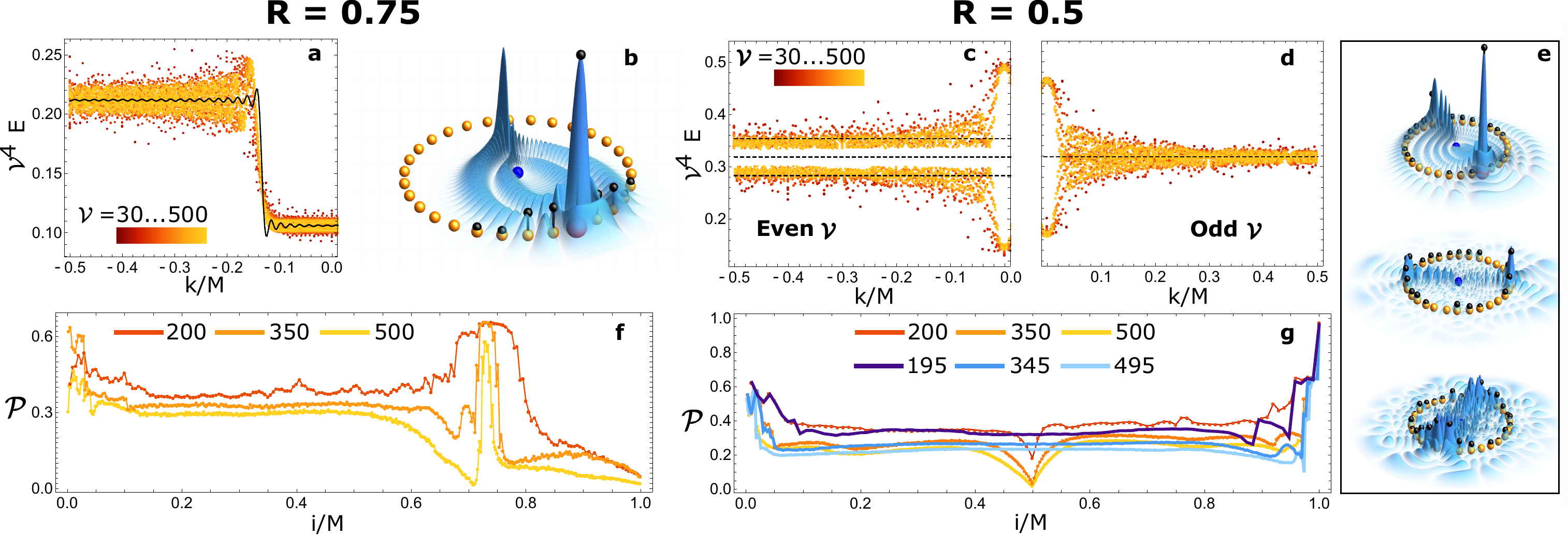}
	\caption{\label{fig:R075} 
 {Characteristics of the perturbed Rydberg atom at  \textit{R} = 0.75 and  \textit{R} = 0.5.}
			(a) Eigenspectra of the $R = 0.75$ ring for $30\le\nu\le500$ plotted as a function of wave number (the mirror-image $k>0$ spectra are not shown). The black curve shows the approximate spectrum obtained in the  $\nu\to\infty$ limit. (b) the $R = 0.75$, $\nu=30$ trilobite state. 
			(c,d) Eigenspectra of the $R = 0.5$ ring. The spectra for even $\nu$ are plotted only for negative $k$ in (c) and odd $\nu$ values are plotted only for positive $k$  in (d). The black lines correspond to the flat band and on-site energies for $\nu=500$ discussed in Appendix \ref{sec:app:matrixel}.
			(e)   The $R = 0.5$ trilobite state for $\nu=30$ and two exemplary eigenstates of the disordered system. 
			(f), (g) $\mathcal{P}$ distributions for several $\nu$ values with fixed radial (for $R = 0.75$) or angular (for $R = 0.5$) disorder.
 	}

\end{figure*}

The energy-resolved participation ratios shown in Fig.~\ref{fig:R1} (c) and (d) provide insight into the role of correlations and the distinction between on- and off-diagonal disorder \cite{titovNonuniversality2005,kuhlEnhancement2008,izrailevAnomalous2012,soukoulisOffdiagonal1981}.
The positively correlated off-diagonal radial disorder manifests itself in the pronounced asymmetry seen in Fig.~\ref{fig:R1} (c), especially in contrast to the \textit{anti}-correlated off-diagonal disorder in Fig.~\ref{fig:R1} (d), where only a small  residual asymmetry is present due  to the negative next-nearest-neighbor hopping term (see Appendix \ref{sec:app:matrixel}).
A sharp feature in the band middle depends on the parity of $M$: when $M$ is odd (even) there is a minimum (maximum).
A state with infinite localization length is predicted to occur at the exact band middle in one-dimensional models with off-diagonal disorder  \cite{soukoulisOffdiagonal1981,theodorouExtended1976,brouwerDensity2000}; this could be the source of this feature, which is further modified by the correlated disorder. 

\subsection{\textit{R} \textless\, 1: long-ranged interactions}

{To
illustrate the diversity of localization scenarios 
possible with a perturbed Rydberg atom, we briefly discuss} two other ring sizes, $R = 0.75$ and $ R = 0.5$. 
As seen in the trilobite states plotted in Fig.~\ref{fig:R075} (b) and (e), the hopping amplitudes for these cases extend over some ($R = 0.75$) and all ($R = 0.5$) sites.  
We will first  contrast the disorder-free properties of these two systems before {discussing} their responses to the presence of disorder. 

For $R = 0.75$, the hopping terms oscillate as a function of $|q - q'|$ before decaying rapidly around $|q - q'|\approx M/10$. 
At $\nu\to\infty$, the continuous form of the hopping amplitudes tends asymptotically toward a sinc function, $V_{qq'}\sim\nu^{-4}\text{sinc}[\pi\sqrt{3} (q-q')]$, as discussed in Appendix \ref{sec:app:matrixel}.  
{As shown in Fig.~\ref{fig:R075}(a)} 
this results in an eigenspectrum closely approximated by a box function,  whose flat bands are broadened by the deviations from the asymptotic form of the hopping ampl{it}udes.
Note that the spectra  are only shown for half the range of allowed wave numbers, {since they are symmetric about $k= 0$}.

On the other hand, the $R = 0.5$ hopping amplitudes oscillate over the entire ring, rising to a maximum at the opposite side (see Fig.~\ref{fig:R075}(e)). 
The {effect is particularly strong}  for even values of $\nu$, leading to a dimerization of the system \cite{phillipsLocalization1991} {and}
strongly impacting the observed disorder-free eigenspectra shown in Fig.~\ref{fig:R075}(c).
These spectra condense into two relatively flat bands separated by a wide band gap when $\nu$ is even, or a single band when $\nu$ is odd. 
We find that the dominant hopping amplitude $V_{q,q+M/2}$  scales as $\nu^{-13/3}$, while the other hopping amplitudes scale as $\nu^{-5}$. 
When $\nu$ is even, the model Hamiltonian discussed in Appendix \ref{sec:app:matrixel} shows that the width of the band gap scales as $\nu^{-1/3}$ and thus closes in the thermodynamic limit. 
The strongly split levels around $k = 0$ are manifestations of the all-to-all coupling, and survive in the thermodynamic limit, as shown for a simplex model \cite{ossipovAnderson2013}. 

We now analyze which phenomena in the disordered cases arise because of these different spectral features and hopping amplitudes. 
Fig.~\ref{fig:R075}(f) shows three $\mathcal{P}$ distributions for the radial-disordered $R = 0.75$ system.
The regions with nearly flat bands localize uniformly. 
The levels lying in the band gap are well-separated in energy, impeding localization, but as $\nu$ increases the gaps between these levels is found numerically to close approximately as $\nu^{-0.27}$.
This causes the band of extended states visible in Fig.~\ref{fig:R075}(f) around $i/M=0.75$ to shrink as $\nu$ increases, suggesting that the boundaries of this region are not mobility edges but rather finite size effects. 
 
 The $\mathcal{P}$ distributions for the angular-disordered $R = 0.5$ system are shown in Fig.~\ref{fig:R075}(g). 
As in the previous cases, localization occurs most rapidly at band edges: the band gap present in the even-$\nu$ spectrum leads to a pronounced valley in the participation ratio that is absent in the odd $\nu$ case. 
 Fig.~\ref{fig:R075}(e) shows two exemplary $\nu=30$ eigenstates from this valley. 
These are approximately symmetric under reflection and localize on two opposite sites due to the dominant opposite-neighbor coupling. 
Although the overall $\mathcal{P}$ distributions shrink to lower values as $\nu$ increases, 
we find that states near $k = 0$, for this disorder strength and range of $\nu$, appear to remain extended. 
This is akin to the behavior of systems with sufficiently long-range power-law interactions, which have an extended state at the band edge in the thermodynamic limit \cite{rodriguezAnderson2003a,demouraLocalization2005a,mirlinTransition1996}. 
However, {these results cannot be applied so simply to the Rydberg system for which} long-range correlation and off-diagonal disorder 
can enhance localization 
\cite{nosovCorrelationinduced2019a}.

\section{Conclusions and outlook}
  By uncovering and exploiting the surprising relationship between the electronic eigenstates of a Rydberg composite and those of a tight-binding Hamiltonian, we have connected two paradigmatic concepts in atomic and condensed matter physics, showing that the Rydberg electron of a hydrogen-like atom can undergo Anderson localization. 
This mapping is contingent on two atypical conditions in a single-particle system: high degeneracy and an infinite spectrum of bound states.  
Bertrand's theorem states that the only central force potentials in which all bound orbits are closed are the Coulomb and harmonic oscillator potentials \cite{bertrand1873theoreme};  quantum mechanically, this implies that these are unique in providing both the requisite degeneracy and infinite spectrum. 
We expect that the states of a quantum harmonic oscillator will localize under similar conditions as discussed here, which may also further elucidate the supersymmetric links between these systems \cite{alankosteleckySupersymmetry1985}. 
The study of the two-dimensional hydrogen atom or elliptical harmonic oscillators could reveal the role of inherent symmetry properties of the underlying structure in the localization properties \cite{keski-rahkonenQuantum2019}. 
A ring of ground state atoms confined to a ring is not the only interesting implementation of a Rydberg composite. 
Two-dimensional systems could be considered by arranging scatterers into a spherical shell, staggered, stacked, or intersecting rings, or a helix. 
More generally, Rydberg atoms can be perturbed by external fields rather than ground state atoms;  dynamical localization and localization in the time domain have both been predicted to occur in microwave-driven Rydberg atoms \cite{giergielAnderson2017a,schelleMicrowaveDriven2009a}.

	We close with a few comments on the experimental study of the isomorphism between these two systems that we have described here. The experimental realization of a Rydberg atom in a designed environment involves tradeoffs between the challenges of preparing and manipulating high Rydberg states and the difficulty of positioning ground state atoms. 
 Although a demonstration of Anderson localization as we have strictly defined it here in the thermodynamic limit would require large $\nu$ and many scatterers, we stress that interesting localization effects can already be seen for small numbers of scatterers and moderate $\nu$.  
		Experimental signatures of the localization length could be provided by observables properties such as the photoionization rate or dipole moments of the eigenstates.
		
		To avoid the challenges associated with trapping the scatterers in tweezer arrays close to the Rydberg atom, one could study instead a Rydberg atom in a dense ultracold gas. 
		Such experiments are routinely performed at densities where many tens or hundreds of atoms are found within the Rydberg orbit \cite{shafferUltracold2018a}.  
   Due to the random positions of the scatterer atoms, the corresponding tight-binding system is characterized by strong on-site disorder and a complicated set of strongly disordered hopping amplitudes.
   Although the phenomenology of localized states under these conditions is generally known \cite{luukkoPolyatomic2017,abumwisExtended2020}, 
			 characterizing localization systematically in the thermodynamic limit will be more challenging here, as the disorder is very strong and uncontrolled and it is not even clear if a thermodynamic limit exists.


	\acknowledgements
		The authors are grateful for numerous valuable discussions with P.~Giannakeas and A. Hunter. M.T.E. and A.E. thank I. Khaymovich for useful discussions regarding long-range hopping. 
			M.T.E acknowledges partial support from the Alexander von Humboldt Stiftung. 
			AE acknowledges support from the DFG via a Heisenberg fellowship (Grant No EI 872/10-1).

\appendix
\section{Additional details of the Hamiltonian}
\label{sec:app:hamiltonian}
In this appendix we show how to construct the Hamiltonian (Eq.~\ref{eq:originalham}) from the microscopic Hamiltonian of the perturbed Rydberg atom,
\begin{equation}
	\label{eq:methodsH}
	H=  H_\text{Ryd}+ \sum_{q=1}^M H_\text{int}(q). 
\end{equation} 
The Hamiltonian for an alkali Rydberg atom is
\begin{equation}
    \label{eq:app:unpertH}
   H_\text{Ryd}= -\frac{\nabla^2}{2}-\frac{1}{ r}+V_\text{sr}(r),
\end{equation}
where
$V_\text{sr}(r)$ is an empirically derived potential parameterizing the effect of the multi-electron core \cite{eilesTrilobites2019}.
Because of this non-Coulombic potential, the eigenenergies of $H_\text{Ryd}$ are in general different from those of hydrogen. 
The energies follow a modified Rydberg formula including a set of energy-independent quantum defects $\mu_l$, i.e.
\begin{equation}
H_\text{Ryd}\ket{\nu i} = -\frac{1}{2(\nu-\mu_{l})^2}\ket{\nu i}.
\end{equation}
We use the Fermi pseudopotential to describe the interaction between the electron and a scatterer: \cite{fermiSopra2008}
\begin{equation}
     H_\text{int}(q)=2\pi a_s[k(R_q)]\delta^3(\vec r - \vec R_q).
 \end{equation}
In doing so, we include only the contribution due to $s$-wave scattering of the electron from each atom. 
This is justified due to the low kinetic energy characteristic of the Rydberg electron, which suppresses the influence of higher order partial waves. 
 
Expressing $H_\text{Ryd}$ in terms of its eigenstates $\ket{\nu i}$ and rewriting the Fermi pseudopotentials yields the Hamiltonian 
\begin{equation}
	\label{eq:sm:rydberghamiltonian}
	H= -\sum_{\nu i}\frac{\ket{\nu i}\bra{\nu i}}{2(\nu-\mu_{l_i})^2}+2\pi \sum_{q=1}^Ma_s[k(R_q)]\ket{\vec R_q}\bra{\vec R_q}.
\end{equation} 
The quantum defects in the first term break the level degeneracy which is crucial for the separable form of $\mathcal{H}$. 
However, because of the short-ranged character of $V_\text{sr}(r)$, $\mu_l\approx 0$ in alkali atoms for all $l> 3$. 
The derivations used throughout can therefore proceed just by excluding the low angular momentum states from the degenerate manifold. 
For simplicity, we have set all $\mu_l = 0$ in our calculations. 
To obtain Eq.~\ref{eq:originalham} all that remains is to express Eq.~\ref{eq:sm:rydberghamiltonian} in the basis of degenerate states with fixed $\nu$.  

The coupling between different degenerate manifolds is already negligible for the smallest $\nu$ we consider here ($\nu=30$) due to the relatively weak effect of the scatterers compared to the overall Coulomb energy scale.  
Furthermore, this coupling will vanish in the thermodynamic limit because the energy separation between Rydberg manifolds, determined by the first term in Eq.~\ref{eq:sm:rydberghamiltonian}, scales as $\nu^{-3}$, while the width and center of mass of the perturbed subspace drop off as $\nu^{-4}$ or faster.
Thus, we are well-justified in considering each degenerate $\nu$-manifold of $H$ separately for all $\nu$ considered. 
If desired, the three approximations discussed here - the neglect of higher order partial waves, of non-perturbative coupling to other $\nu$ manifolds, and of quantum defects - can be relaxed using the generalized trilobite orbital protocol developed in Ref.~\cite{eilesTrilobites2019}.

\section{Transformation between Rydberg atom and lattice representations}
\label{sec:app:transform}
In this appendix, we discuss in more detail the transformation $\hat U$ used in the text and the transformation between eigenstates in both representations.  We use Einstein notation when summing over repeated indices, using $q$ or $p$ to label scatterer indices which run from  $1$ to $M$ and indices $i$ and $j$ to label Rydberg basis states, which range from $1$ to $\nu^2$.  
For all scenarios considered in this article, $\nu^2>M$. 
To review the relationships defined in the main text, we have
\begin{align}
    \mathcal{H}_{ii'} &= 2\pi\mathcal{W}_{iq}\mathcal{W}^\dagger_{qi'}\\
    U_{qi} &= \left[(\mathcal{W}_{}^\dagger\mathcal{W}_{})^{-1/2}\right]_{qp}\mathcal{W}_{pi}^\dagger\label{eq:sm:Udef}\\
    U_{jq}^\dagger &= \mathcal{W}_{ip}\left[(\mathcal{W}_{}^\dagger\mathcal{W}_{})^{-1/2}\right]_{pq}.
\end{align}
It is straightforward to show from these definitions that $U$ is a semi-unitary transformation satisfying
\begin{align}
    U_{qi}U_{iq'}^\dagger &= \delta_{qq'}\\
    U_{iq}^\dagger U_{qi'}  &=\mathcal{W}_{ip}\left[(\mathcal{W}^\dagger\mathcal{W})^{-1}\right]_{pq}\mathcal{W}_{qi'}^\dagger=Q_{ii'}.
    \end{align}
    Furthermore, using these derivations, we have
    \begin{align}
        \mathcal{H}_{ii'} &= Q_{ij}\mathcal{H}_{jj'}Q_{j'i'}\label{eq:app:Q}\\
    H_{qq'} &= U_{qi}\mathcal{H}_{ii'}U_{i'q'}^\dagger.
\end{align}
We use Eq.~\ref{eq:app:Q} to rewrite the eigenvalue equation yielding the eigenenergies in the Rydberg basis:
\begin{align}
	\label{eq:sm:hamtrilobite}
\nonumber	Ev_{i}&=\mathcal{H}_{ii'}v_{i'} \\
\nonumber	&= U^\dagger_{iq}U_{qj}\mathcal{H}_{jk}U^\dagger_{kp}U_{pi'}v_{i'}.
\end{align}
Applying $U_{qi}$ to both sides and defining the transformed eigenvector
\begin{equation}
    \label{eq:sm:transformvec}
    \tilde v_q = U_{qi}v_i
\end{equation}
gives
\begin{align} E\tilde v_{q'}	=U_{qj}\mathcal{H}_{jk}U_{kq'}^\dagger\tilde v_{q'} = H_{qq'}\tilde v_{q'}.\end{align}
Diagonalization of $H$ yields the $M$ non-zero eigenvalues $E_k$ of $\mathcal{H}$ and their associated eigenvectors $\tilde v_q^{(k)}$. 

To complete this section, we show the transformation between these eigenvectors and those in the Rydberg basis, $v_i^{(k)}$. Using Eq.~\ref{eq:sm:Udef} to rewrite Eq.~\ref{eq:sm:transformvec} yields
\begin{align}
\left[(\mathcal{W}_{}^\dagger\mathcal{W}_{})^{1/2}\right]_{qp}\tilde v_{p}^{(k)} &= \mathcal{W}_{qi}^\dagger v_i^{(k)}.
\end{align}
Since $\tilde v_p^{(k)}$ is an eigenvector of $H$, we can replace the matrix on the left side of this equation with the eigenvalue $E_k$:
\begin{align} \sqrt{E_k/(2\pi)}\tilde v_q ^{(k)}&= \mathcal{W}_{qi}^\dagger v_i^{(k)}.\end{align}
Multiplying by $2\pi \mathcal{W}_{i'q}$ leads to 
\begin{align}
	 \sqrt{2\pi E_k}\mathcal{W}_{i'q}\tilde v_q^{(k)} &= \mathcal{H}_{i'i}v_{i}^{(k)}.
\end{align}
Using the eigenvalue equation again to replace the right hand side of this expression with $E_kv_i^{(k)}$ yields the relationship 
\begin{equation}
\label{eq:sm:evecs}
    v_i^{(k)} = \sqrt{\frac{2\pi}{E_{k}}}\mathcal{W}_{iq}\tilde v_q^{(k)}
\end{equation}
between the eigenvectors in the two representations.

\section{Equivalence of localization measures}
\label{sec:app:loca}
We have used the normalized participation ratio $\mathcal{P}$ to characterize the localization of the electron and relate its localization behavior to that of a particle in a lattice. 
As defined in Eq. \ref{eq:NPRk}, $\mathcal{P}$ characterizes localization in the site basis.
However, 
a desirable condition is that localization in this representation remains physically meaningful when we discuss the localization of the Rydberg electron. 
As seen in the eigenstate figures presented in the text and the relationship Eq.~\ref{eq:sm:evecs}, the amplitudes of the eigenstates in both representations are clearly related; here we make this argument rigorous by showing that the normalized participation ratio computed in position space is equivalent to what is calculated in the site basis. 

To begin, we compute the probability of finding the electron at the position of scatterer $p$,
\be
\text{Prob}(p)=\left|\sum_iv_i^{(k)}\phi_i(\vec R_p)\right|^2,
\ee
which can be rewritten in terms of $\mathcal{W}$,
\be
\text{Prob}(p) = \left|\sum_i\mathcal{W}_{pi}^\dagger v_i^{(k)}\right|^2.
\ee
By then transforming the eigenvector $v_i^{(k)}$ into the site basis we obtain
\begin{align}
\label{eq:app:probp}
\text{Prob}(p)&= \left|\sum_{i,q}\mathcal{W}^\dagger_{pi}\frac{\mathcal{W}_{iq}\tilde v_q^{(k)}}{[\epsilon^{(k)}]^{1/2}}\right|^2 =|\epsilon^{(k)}|\left|\tilde v_p^{(k)}\right|^2,
\end{align}
where we have recognized the appearance of the Hamiltonian matrix acting on the eigenvector. 

We define $\mathcal{P}_{spatial}(k)$, the normalized spatial participation ratio, by considering the probabilities to find the electron at any of the scatterer positions,
\begin{align}
\label{eq:app:Pspat}
\mathcal{P}_{spatial}(k) &= \left(M\sum_P\left|\text{Prob}(p)\right|^2\right)^{-1}.
\end{align}
Using Eq.~\ref{eq:app:probp} in Eq.~\ref{eq:app:Pspat} and comparing the result with the definition of $\mathcal{P}(k)$, we see that $\mathcal{P}_{spatial}(k) = \mathcal{P}(k)/[\epsilon^{(k)}]^{2}$. Localization in the site basis implies spatial localization, albeit with a normalization factor given by the eigenenergy of state $k$. 

This normalization factor can be removed by considering relative spatial probabilities in the formulation of $\mathcal{P}_{spatial}(k)$, since the most relevant localization measure is not localization relative to the entire allowed volume (which our previous measure characterizes) but instead localization within the spatial volume of interest.
The probability of finding the electron at the position of one scatterer relative to the total probability of finding it at any scatterer is
\be
P(p) = \frac{\text{Prob}(p)}{\sum_P\text{Prob}(p)}=|\tilde v_p^{(k)}|^2.
\ee
Note that, as long as the eigenvectors are normalized, this probability is normalized so that there is unit probability to find the electron on the ring of scatterers, i.e. $\sum_p P(p)=1$. 
Using this probability to define the spatial participation ratio gives
\be
\mathcal{P}_{spatial}(k) = \left(M\sum _p\left|P(p)\right|^2\right)^{-1}=\left(M\sum_p|\tilde v_p^{(k)}|^4\right)^{-1}.\nonumber
\ee 
This final step is identical to $\mathcal{P}(k)$, and thus we conclude that the two participation ratios are equivalent.

\section{Numerical details of the disorder}
\label{sec:app:disorder}
We introduce disorder to the ring Rydberg composite either by shifting the angles of the scatterers, $\phi_q \to \frac{2\pi }{M}\left[q+\delta_q(\nu)\right]$, or their positions, $R_q \to [1+\overline\delta_q(\nu)]R_q$. 
We take $\delta_q,\overline\delta_q$ to be independent Gaussian random variables with variance $\sigma^2$ and mean zero. 
For the specific examples shown in the  text, we choose different values of $\sigma$. For $R = 1$, we choose $\sigma=17\times 10^{-3}$ for angle disorder and $\sigma =(2\times 10^{-3})\cdot(30^{2/3})\approx 0.01931$ for radial disorder. 
For $R = 0.75$ we show only results for radial disorder, where $\sigma = 1.33\times 10^{-3}$, and for $R = 0.5$ we show only angular disorder with $\sigma = 22\times 10^{-3}$. 
In all cases we averaged over 1000 disorder realizations.

\section{Matrix elements in the thermodynamic limit}
\label{sec:app:matrixel}
In order to study the behavior of the ring Rydberg composite in the thermodynamic limit, we need to carefully examine the analytical expressions for the matrix elements $H_{qq'}$.
The on-site potentials $E_q$ and hopping amplitudes $V_{qq'}$  are  \cite{eilesRing2020,eilesTrilobites2019}
\begin{align}
	\label{eq:sm:energydefs} 
	E_q   &=\sum_{lm}\mathcal{R}_{lm}(R_q,R_q)\\
	V_{qq'}&= \sum_{lm}\mathcal{R}_{lm}(R_q,R_{q'})e^{-im(\phi_q - \phi_{q'})},
\end{align}
where
\begin{align}
\label{eq:rdef}
\mathcal{R}_{lm}(R_q,R_{q'}) &= \nonumber\frac{(l+\frac{1}{2})(l-m)!(l+m)!}{\left[\left(\frac{l+m}{2}\right)!\left(2^{(l+1)}\frac{l-m}{2}\right)!\right]^2}\\&\times \left[\frac{u_{\nu l}(2\nu^2R_q)}{R_q\nu^2}\right]\left[\frac{u_{\nu l}(2\nu^2R_{q'})}{R_{q'}\nu^2}\right].
\end{align}
 The functions $u_{\nu l}(r)$ are the reduced hydrogen radial functions and $u'_{\nu l}(r)=\frac{d u_{\nu l}(r)}{dr}$. 
These expressions can be analytically summed \cite{eilesTrilobites2019}, yielding
\begin{align} 
	\label{eq:energydefs1} 
	E_q   &=\frac{(R_q^{-1}- 1)[u_{\nu 0}(2\nu^2R_q)]^2 + \nu^2[u_{\nu 0}'(2\nu^2R_q)]^2}{2\nu^2}
  \end{align}
 and
\begin{align}
	\label{eq:energydefs2} 
	V_{qq'}&=  \frac{u_{\nu 0}'(t_-)u_{\nu 0}(t_+) - u_{\nu 0}(t_-)u_{\nu 0}'(t_+)}{2(t_+-t_-)},
\end{align}
with
\begin{align}
	\label{eq:rdef}
	t_\pm &= \nu^2\left(R_{q}+R_{q'}\right)\\&\pm \nu^2\sqrt{R_q^2+R_{q'}^2-2R_qR_{q'}\cos\left(\phi_q - \phi_{q'}\right)}\nonumber. 
\end{align}
It is remarkable that only the $s$-wave radial wave function is necessary to evaluate these expressions, making them useful for determining asymptotic properties and for computations.
Although $u_{\nu0}(r)$ and $u_{\nu 0}'(r)$ oscillate as a function of $\nu$, some well-defined limits exist:
\begin{align}
	\label{eq:sm:zedgeu}
 u_{\nu0}(0)&= 0, \\
	\label{eq:sm:zedgeup}
	u'_{\nu 0}(0)&=2\nu^{-3/2},\\
	\label{eq:sm:edgeu}
	\lim_{\nu\to\infty}u_{\nu0}(2\nu^2)&= a\nu^{-5/6}\\
	\label{eq:sm:edgeup}
	\lim_{\nu\to\infty}u'_{\nu 0}(2\nu^2)&=b{\nu^{-13/6}},
\end{align}
where $a\approx-0.56355$ and $b \approx 0.326$. 

We can study the behavior of the composite for the three $R$ values discussed in the text. 
For each, we will study the asymptotic behavior of the matrix elements in the disorder-free scenario to uncover their scaling behavior and develop useful model Hamiltonian. 
Following that, we will treat the influence of disorder analytically by expanding the Hamiltonian matrix elements to first order in the positional disorder, and perform the asymptotic analysis again to find the influence of positional disorder on the energies. 

$\mathbf{R=1}$ \textbf{ring:} The $R = 1$ case has some particular scaling laws not followed by the other ring sizes due to its size coinciding with the outermost classical turning point.  
From Eqs.~\ref{eq:energydefs1},~\ref{eq:energydefs2}, and \ref{eq:sm:edgeup}, we obtain
\be
\lim_{\nu\to\infty}E_{q} = \frac{[u'_{\nu 0}(2\nu^2)]^2}{2}\approx \frac{b^2\nu^{-13/3}}{2}.
\ee
Plugging in numerical values yields $E_{q}=a_1\nu^{-13/3}$ where $a_1\approx 0.053138$.
To obtain off-diagonal matrix elements with the same $\nu^{-13/3}$ scaling, we find that we must set $M=\text{Floor}(3\nu^{2/3})$.
With this choice, the largest hopping amplitudes are 
\begin{align}
	\lim_{\nu\to\infty}V_{qq+1} &=b_1{\nu^{-13/3}},\,\,b_1\approx {0.01355}\\
	\lim_{\nu\to\infty}V_{qq+2} & =-c_1\nu^{-13/3},\,\,c_1\approx {0.0004}.
\end{align}
We see that the next-nearest neighbor hopping amplitude is already thirty times smaller than the nearest-neighbor amplitude;
longer-ranged hopping amplitudes continue to decrease in size and are negligible. 
This justifies the model Hamiltonian discussed in the text,
\begin{align}
    H_\text{model}^{R=1} \nonumber= \nu^{-13/3}\sum_{q=1}^M &\Bigg(a_1\ket{q}\bra{q} + b_1 \ket{q}\bra{q+1}\\& -c_1\ket{q}\bra{q+2}+\text{c.c}\Bigg).
\end{align}
The next-nearest-neighbor hopping term creates the slight asymmetry in the eigenspectrum shown in Fig. \ref{fig:R1}.

We now perturb the positions of the scatterers as prescribed in Appendix \ref{sec:app:disorder}. The matrix elements are, for large $\nu$, 
\begin{align}
	\nu^{13/3}E_{q}&\approx a_1-g_1\nu^{2/3}\delta_q\\
	\nu^{13/3}V_{qq+1}&\approx b_1-f_1(\overline\delta_{q'} - \overline\delta_q)-e_1\nu^{2/3}(\delta_q + \delta_{q'})\\
	\nu^{13/3}V_{qq+2}&\approx -c_1,
\end{align}
where $e_1\approx0.015$, $f_1 \approx0.04$, and $g_1 \approx0.1519$. 
As stated in the text, angle disorder (terms depending on $\overline{\delta_q}$) leads to anti-correlated off-diagonal disorder in the Hamiltonian matrix elements, proportional to $\delta_q - \delta_{q'}$. 
On the other hand, radial disorder leads to uncorrelated diagonal disorder proportional to $\delta_q$ and a weaker, correlated off-diagonal disorder  proportional to $\overline\delta_q + \overline{\delta_{q'}}$. 
These conclusions are true in general for all values of $R$. 
However, a special feature of this $R = 1$ case is the scaling of the terms proportional to $\delta_q$ with $\nu^{2/3}$. 
This indicates that we must rescale the radial \textit{positional} disorder by a factor $\nu^{-2/3}$ in order to provide a constant \textit{matrix element} disorder as $\nu$ increases in order to obtain a proper thermodynamic limit.

$\mathbf{R=0.75}$ \textbf{ring:} 
This $R$ value is more challenging to address analytically as there are more non-zero hopping elements to consider than in the previous case. 
The diagonal matrix elements are
\be
E_{q} = \frac{1}{2\nu^2}\left[\frac{1}{3}(u_{\nu 0}(3/2\nu^2))^2+\nu^2(u'_{\nu 0}(3/2\nu^2)\right]^2,
\ee
which simplifies further as $\nu\to\infty$ to become $E_{q}\approx \frac{a_{0.75}}{\nu^4}$ with $a_{0.75} =0.1833$. 
In this same limit, the hopping terms for relatively small $|q - q'|<\nu/20$, approach the functional form
\be
\label{sup:eq:R075int}
V_{qq'} \sim \frac{a_{0.75}}{\nu^4}\frac{\sin[\omega (q-q')]}{\omega  (q - q')},
\ee
where $\omega = \pi\sqrt{3}$. For larger distances along the ring, $\nu/20 < |q - q'| < \nu/10$ (approximately), the hopping terms continue to oscillate, but with mostly constant amplitude. 
At $|q - q'| \approx \nu/10$, the hopping terms rapidly decay to zero. 

As in the $R = 1$ case, we use this as the basis for an analytic model for this case, although it is gives only qualitative insight here. 
We assume that the hopping amplitudes follow Eq.~\ref{sup:eq:R075int} for $|q-q'|<\nu/10$ and vanish for all other $q,q'$ pairs. 
The eigenspectrum of such a Hamiltonian, shown as the black curve in Fig.~\ref{fig:R075}(a), is a very good qualitative match to the actual Rydberg spectrum. 
The cutoff length, after which the hopping amplitudes vanish, affects the bandwidth of the two individual bands, which flatten and narrow as the cutoff length increases.

Performing the same disorder analysis as for $R = 1$, we find that the energy disorder stemming from radial disorder is proportional to $\delta\nu^{-4}$, and therefore the radial disorder in this case needs no additional scaling.  
The angular disorder scaling behaves similarly as before, requiring no additional $\nu$ scaling beyond what is done for the matrix elements. 

\begin{figure}[t]
	
	\includegraphics[width=\columnwidth]{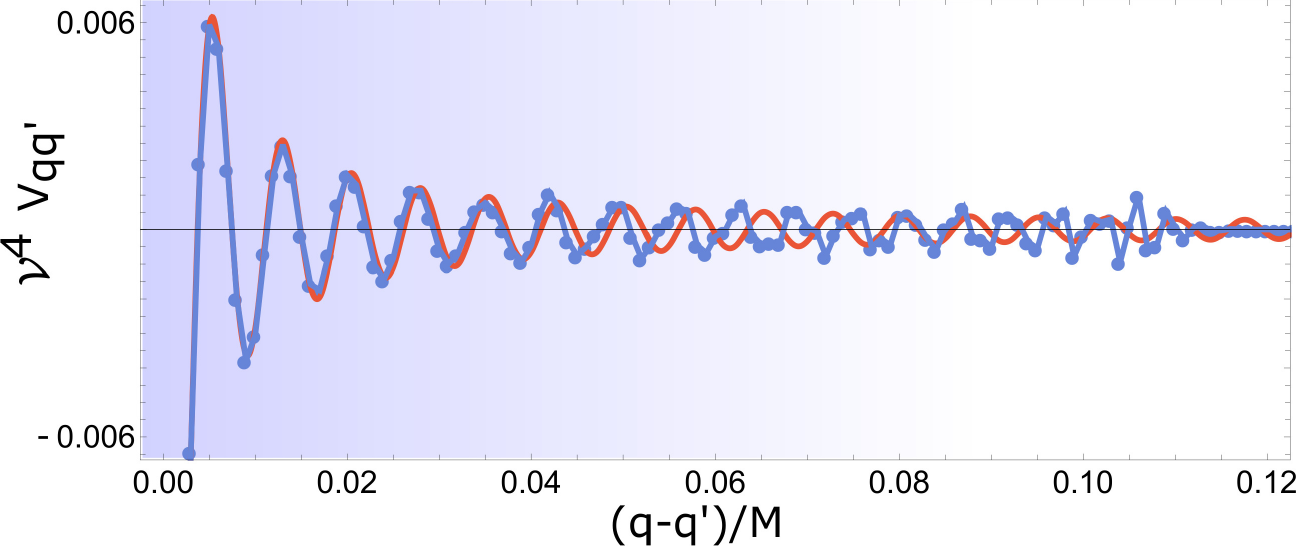}
	\caption{\label{fig:int} {Exact hopping amplitudes for $\nu = 1000$ for the $R = 0.75$ ring (blue) compared with the asymptotic form, Eq.~\ref{sup:eq:R075int} (red). The points mark specific scatterer positions. } 
	}
\end{figure}

$\mathbf{R=0.5}$ \textbf{ring:} 
The diagonal elements are straightforward to evaluate for the $R = 0.5$ ring: 
\be
E_{q} = \frac{1}{2\nu^2}\left([u_{\nu 0}(\nu^2)]^2+\nu^2[u'_{\nu 0}(\nu^2)]^2\right),
\ee
which becomes $E_{q}\approx a_{0.5}\nu^{-4}$, 
where $a_{0.5}\approx 0.3183$, as $\nu\to\infty$. 
The largest hopping amplitude, due to the shape of the trilobite orbitals, connects  site $q$ to site $q + M/2$ if $M$ is even and site $q+(M\pm 1)/2$ if $M$ is odd. 
The parity of $M$ therefore plays a key role in overall form of the eigenspectrum, in contrast to the previous $R$ values where it was irrelevant. 
When $M$ is even, the dominant hopping term is
\be
\label{eq:smR05hopping}
V_{q,q+M/2}=\frac{u_{\nu 0}'(0)u_{\nu 0}(2\nu^2)}{4\nu^2} = \frac{u_{\nu 0}(2\nu^2)}{2\nu^{7/2}}.
\ee 
Asymptotically, this scales like $V_{q,q+M/2}\approx-c_{0.5}\nu^{-13/3}$, where $c_{0.5}\approx 0.2818$. 
When $M$ is odd, the two neighboring particles on the opposite side of the ring have identical amplitudes, also scaling like $\nu^{-13/3}$. 
For other hopping terms the expression $V_{qq'}$ no longer depends on the exceptional scaling of $u_{\nu 0}(2\nu^2)$ and $u'_{\nu0}(2\nu^2)$ (see Eqs.~\ref{eq:sm:edgeu}-\ref{eq:sm:edgeup}) and we can use the "standard" scaling behavior $u_{\nu 0}(\nu^2) \sim \nu^{-1}$ and $u'_{\nu 0}(\nu^2)\sim \nu^{-2}$. 
Together this gives a $\nu^{-5}$ scaling for the off-diagonal elements. 

This ring is therefore distinct from the other two that we have considered in that its matrix elements scale with $\nu$ in different ways. 
To understand the resulting eigenspectrum in the even parity case, we construct an approximate Hamiltonian, specializing to the even $M$ case for simplicity. 
This approximate Hamiltonian is
\begin{align}
    H_\text{model}^{R=0.5}=\nu^{-4}\sum_{q = 1}^M&\Bigg[a_{0.5}\ket{q}\bra{q} -c_{0.5}\nu^{-1/3}\ket{q}\bra{q+M/2}\nonumber\\& + \sum_{q'\ne q, q+M/2}b_{0.5}\nu^{-1}\ket{q}\bra{q'}\Bigg]\label{eq:app:modelH05}.
\end{align}
The crucial simplification in this Hamiltonian that is not present in the exact Rydberg Hamiltonian is the assumption of a constant hopping amplitude for all sites except the one on the opposite side of the ring. 
In actuality, the amplitudes vary over the entire range of the ring both in sign and amplitude. 
We ignore the overall $\nu^{-4}$ scaling  trivial diagonal energy $a_{0.5}$ in the following. 
Eq.~\ref{eq:app:modelH05} can be diagonalized analytically to gain some insight into the Rydberg composite's spectrum.
It has a $\nu/2$-degenerate eigenvalue $e_1=c_{0.5}\nu^{-1/3}$, a $\nu/2-1$ degenerate eigenvalue $e_2=-2b_{0.5}\nu^{-1}-c_{0.5}\nu^{-1/3}$, and a single eigenvalue $e_3=b_{0.5}(\nu-2)\nu^{-1}-c_{0.5}\nu^{-1/3}$. 
For large $\nu$ the eigenspectrum consists of two flat bands at the energies $\pm c_{0.5}\nu^{-1/3}$, and a single state lying at $b_{0.5}-c_{0.5}\nu^{-1/3}$. 
In the thermodynamic limit the band gap closes completely and the system condenses to a flat band at zero energy and a single shifted state at energy $b_{0.5}$. 
In this limit, the system qualitatively resembles the odd-parity $M$ state. 
Although this is a highly simplified qualitative picture of the $R = 0.5$ Rydberg eigenspectra, the basic features exist also in the real case, as can be seen by studying Fig.~\ref{fig:R075}c. 

Just as the disorder-free properties of this ring size are more complicated than those of the other rings, so is its disorder scaling. 
Angle disorder perturbs the dominant hopping amplitude by a term which is \textit{second order} in the positional disorder strength, but overall scales as $\nu^{-13/3}$, the same scaling as the disorder-free value.  
Angle disorder is, therefore, weaker in this case than in the previous $R$ values. 
Under radial disorder, this hopping amplitude is perturbed by a first-order term in the positional disorder strength. 
This term scales as $\nu^{-11/3}$, which, like in the $R = 1$ case, demands that the positional disorder be rescaled to obtain a consistent thermodynamic limit scaling. 
Like the $R = 1$ case, this also requires a rescaling of the positional disorder, $\overline\delta \to \nu^{-2/3}\overline\delta$. 
This rescaling would cause the diagonal disorder to vanish as $\nu\to\infty$, as it scales with disorder as $\nu^{-4}$. 

Since all of the relevant matrix elements discussed in this section decrease as a function of $\nu$ as either $\nu^{-4}$ or $\nu^{-13/3}$, the relevant timescales for the dynamics in the tight-binding Hamiltonian grow as $\nu^4$ or $\nu^{13/3}$, respectively. The lifetime of the Rydberg atom, on the other hand, increases (averaging over all $l$ and $m$ states of a fixed $\nu$) as $\nu^{9/2}$ \cite{gallagherRydberg2005}. Therefore, as $\nu$ increases, the Rydberg lifetime is guaranteed to be sufficient for interesting dynamics to occur.

\bibliographystyle{apsrev4-1}

\end{document}